\documentclass[a4paper,12pt]{article}

\usepackage [T1]{fontenc}
\usepackage {calligra}
\usepackage [T1]{pbsi}
\usepackage{CJKutf8} 
\usepackage[colorlinks=true,linkcolor=blue]{hyperref}
\usepackage{amsmath,amssymb,amsfonts} 
\usepackage{graphicx}                 
\usepackage{color}                    
\usepackage{hyperref}                 
\usepackage[left=1cm,top=1cm,bottom=1cm,right=1cm,nohead,nofoot]{geometry} 
\definecolor{red}{rgb}{1,0,0}           
\definecolor{green}{rgb}{0,1,0}
\definecolor{blue}{rgb}{0,0,1}
\definecolor{darkblue}{rgb}{0,0,0.5}
\definecolor{lightblue}{rgb}{.5,.5,1}
\definecolor{lightgray}{gray}{.87}          
\definecolor{Dark}{gray}{.20}
\definecolor{pink}{rgb}{.95,0.82,0.92}  
\definecolor{yellow}{rgb}{1,1,0}
\definecolor{lightyellow}{rgb}{1,1,.5}
\definecolor{purple}{rgb}{0.7,0,0.85}
\definecolor{darkgreen}{rgb}{0,0.5,0}
\definecolor{orange}{rgb}{0.8,0.2,0.2}
\def \be {\begin{equation}}
\def \ee {\end{equation}}
\def \bea {\begin{eqnarray}}
\def \eea {\end{eqnarray}}
\def \nn {\nonumber}

\def \rr {\raise.35ex\hbox{\small $\prime$}\kern-.17em{\mbox{\large $\imath$}}}
\def \del {\partial}
\def \dels {\partial\kern-.5em / \kern.5em}
\def \As {{A\kern-.5em / \kern.5em}}
\def \Ds {D\kern-.7em / \kern.5em}

\def \d {\delta}
\def \eps {\epsilon}

\def \lam {\lambda}
\def \Lam {\Lambda}

\def \II {I\hspace{-.1em}I\hspace{.1em}}

\def \IIA {\mbox{\II A\hspace{.2em}}}

\def \z {(0)}
\def \nz {(\mbox{\tiny KK})}

\newcommand{\solution}[1]{}

\begin{document}

\pagestyle{plain}

\setcounter{footnote}{0}
\setcounter{section}{0}

\begin{CJK}{UTF8}{bsmi}

\begin{titlepage}

\begin{center}
\hfill UT-11-11
\vskip .5in

\textbf{\LARGE A Non-Abelian Self-Dual Gauge Theory\\
\vskip.2in
in 5+1 Dimensions}

\vskip .5in
{\large
Pei-Ming Ho$^\dagger$\footnote{
e-mail address: pmho@phys.ntu.edu.tw},
Kuo-Wei Huang$^\dagger$\footnote{
e-mail address: kwhuang87@gmail.com},
Yutaka Matsuo$^\ddagger$\footnote{
e-mail address:
matsuo@phys.s.u-tokyo.ac.jp}}\\
\vskip 3mm
{\it\large
$^\dagger$
Department of Physics and Center for Theoretical Sciences, \\
National Taiwan University, Taipei 10617, Taiwan,
R.O.C.}\\
\vskip 3mm
{\it\large
$^\ddagger$
Department of Physics, Faculty of Science, University of Tokyo,\\
Hongo 7-3-1, Bunkyo-ku, Tokyo 113-0033, Japan\\
\noindent{ \smallskip }\\
}
\vspace{60pt}
\end{center}
\begin{abstract}

We construct a non-Abelian gauge theory of chiral 2-forms
(self-dual gauge fields) in 6 dimensions
with a spatial direction compactified on a circle of radius $R$.
It has the following two properties.
(1) It reduces to the Yang-Mills theory in 5 dimensions for small $R$.
(2) It is equivalent to the Lorentz-invariant theory of Abelian chiral 2-forms
when the gauge group is Abelian.
Previous no-go theorems prohibiting non-Abelian deformations of
the chiral 2-form gauge theory are circumvented by
introducing nonlocality along the compactified dimension.

\end{abstract}

\end{titlepage}

\section{Introduction}

The generalization of Abelian gauge theories with 1-form potentials
to higher form potentials is straightforward.
For a $p$-form potential $A^{(p)}$,
we define its gauge transformation by
\be
\d A^{(p)} = d\Lambda^{(p-1)},
\ee
where the gauge transformation parameter $\Lambda^{(p-1)}$
is a $(p-1)$-form,
and the field strength defined by
\be
F^{(p+1)} = dA^{(p)}
\ee
is an invariant $(p+1)$-form.
However, the generalization of higher form potentials $A^{(p)}$ $(p > 1)$
to non-Abelian gauge theories have been a tough challenge to
both theoretical physicists and mathematicians.

In this paper we attack this problem for the case $p=2$,
with the goal of describing the system of multiple M5-branes.
The multiple M5-brane system has been the most challenging
and mysterious brane system in string theory and M theory
\cite{mM5}.
(For a review of M theory branes, see \cite{Berman-review}.)
The salient nature of the M5-brane theory is that it contains
a self-dual gauge field
(also called a {\em chiral 2-form} potential).
Some believed that a Lagrangian formulation for the self-dual gauge theory
was impossible,
because the self-duality condition imposes first order differential equations
on the gauge potentials,
while an ordinary kinetic term
$(\del_{\mu} B_{\nu\lam})(\del^{\mu}B^{\nu\lam})$
always leads to a 2nd order differential equation.
It turns out that the trick is to avoid using some of the components
of the gauge potential,
so that even though we get 2nd order differential equations
from varying the action,
the self-duality condition appears only after integrating once
the equations of motion.
Those components which do not appear in the action appear
as integration ``contants''.
\footnote{
This trick was later generalized in \cite{Chen:2010jgb}
so that for a given spacetime dimension $D$,
one can write down a Lagrangian for
the self-dual gauge field for an arbitrary division of $D$
into two positive integers $D'$ and $D''$ ($D' + D'' = D$).
We refer to it as the $(D' + D'')$-formulation of
the self-dual gauge theory.
}
Hence the Lagrangian for a single M5-brane was first constructed
without manifest Lorentz symmetry,
and a Lorentz-covariant version is possible only by
introducing an auxiliary field \cite{PST,oldM5}.

The gauge symmetry for a single M5-brane in the trivial background is Abelian.
The first non-Abelian gauge theory for self-dual 2-form potentials
was found for an M5-brane in a large $C$-field background \cite{NPM5}.
A double dimension reduction of this M5-brane theory,
called NP M5-brane theory,
\footnote{
``NP'' stands for ``Nambu-Poisson''.
A Nambu-Poisson structure is used to define the non-Abelian gauge symmetry
for the 2-form potential on the M5-brane.
The physical origin of the Nambu-Structure is the coupling of open membranes
to the $C$-field background \cite{ToyModel}
The NP M5-brane theory was first derived from the BLG model \cite{BLG}.
Its gauge field content was further explored in \cite{Pasti,Furuuchi}.
For a brief review, see \cite{HoM5}.
}
is in agreement to the lowest order with
the noncommutative D4-brane action in large NS-NS $B$ field background \cite{NPM5}.
If the NP M5-brane theory can be deformed such that
it agrees with the noncommutative D4-brane theory to all orders,
it would resemble the multiple M5-brane theory.
However, it turns out that it is extremely hard
to deform the NP M5-brane theory \cite{Chen:2010ny}.
We conclude that it takes brand new ideas to construct the multiple M5-brane theory.

In the literature,
there has been various attempts to construct
a non-Abelian gauge theory for a 2-form potential $B$
taking values in a Lie algebra, 
and the corresponding geometrical structures are called ``non-abelian gerbes".
The immediate problem to construct such a model
is that we need to define covariant derivatives $D_{\mu}$,
which need to be specified by a 1-form potential $A$.
For example, in \cite{AB},
the gauge transformations of $A$ and $B$ are defined by
\bea
A' &=& gAg^{-1}+gdg^{-1} + \Lambda, \label{nag1}\\
B' &=& gBg^{-1}+\left[ A',\Lambda\right]_{\wedge} + d\Lambda+\Lambda\wedge\Lambda,
\label{nag2}
\eea
where $g\in G$ is the gauge parameter and $\Lambda\in g$ is a 1-form.
Mathematically such gauge transformations are well-defined,
and suitable to describe some system such as the non-Abelian
generalization of the BF model \cite{BF}.
It is, however, not clear if it is relevant to describe multiple M5.
Physically, the introduction of $A$ increases
the physical degrees of freedom of the system.
For the M5-brane system,
there is no physical degree of freedom corresponding to $A$.
Furthermore, with the addition of $A$,
the field $B$ is not a genuine 2-form potential in the sense that
we can gauge away $A$ by $\Lambda$,
and then $B$ is not independent of its longitudinal components.
The result is similar to
spontaneous symmetry breaking.

An independent attempt to construct non-Abelian 2-form gauge theory
is to define it on the loop space \cite{loopspace}.
This approach introduces infinitely many more degrees of freedom
to the usual Abelian chiral gauge theory even in the Abelian limit.
Instead, our goal is to have a non-Abelian gauge symmetry
which includes the Abelian theory as the special case
when the Lie algebra involved is Abelian.
This criterium is not matched by any existing construction
in the literature.

While a consistent algebra of non-Abelian gauge transformations for
a higher form gauge theory is already difficult to get,
an action for a chiral gauge boson is even more difficult.
Assuming the existence of an action and gauge transformation algebra,
a no-go theorem \cite{no-go}
states that there is no nontrivial deformation of
the Abelian 2-form gauge theory.
One of their assumptions was locality for the action and the gauge transformation laws.
In particular, Lorentz symmetry was {\em not} assumed.

The non-existence of the local action for multiple M5-branes
was argued in another way by Witten \cite{Witten}.
The M5-brane system is known to have conformal symmetry,
which implies that upon double dimension reduction,
the 4+1 dimensional action should be proportional to
\be
\int d^5 x \; \frac{1}{R}.
\label{1/R}
\ee
On the other hand,
the reduction of a 5+1 dimensional local action on a circle should give
\be
\int d^6 x = \int d^5 x \; 2\pi R,
\label{R}
\ee
which has the opposite dependence on $R$.
As long as we assume a Lorentz-covariant formulation for M5-branes
without explicit reference to the compactifiation radius $R$ 
except through the measure of integration,
this gives a strong argument against
the Lagrangian formulation of multiple M5.

Recently there are proposals \cite{Douglas,LPS} claiming that
the multiple M5-brane system compactified on a circle of finite radius $R$
is described by the $U(N)$ super Yang-Mills theory for $N$ D4-branes
even before taking the small $R$ limit.
This would be a duality between two theories in 5 and 6 dimensions, respectively,
but it can not be viewed as an example of the holographic principle of quantum gravity,
because there is no gravitational force in these theories.
Their proposal, if correct, would be revolutionary.
However we will point out its difficulties in Sec.\;\ref{proposal}.

These developments suggest that
it is already a tremendous progress to have a theory for multiple M5-branes
compactified on a circle of finite radius $R$,
if the following two criteria are satisfied:
\begin{enumerate}
\item
In the limit $R\rightarrow 0$,
the theory should be approximated by the gauge field sector of 
the multiple D4-brane theory,
which is $U(N)$ Yang-Mills theory in 5 dimensions.
\item
When the Lie algebra of the gauge symmetry is Abelian,
the theory reduces to $N$ copies of the theory of
6 dimensional Abelian self-dual gauge field.
\end{enumerate}
In view of the no-go theorem \cite{no-go},
the absence of 6 dimensional Lorentz symmetry
due to compactification of the 5-th direction
does not necessarily make the task much easier.
On the other hand,
the 2nd criterium ensures the 6 dimensional Lorentz symmetry
in the broken phase in the limit $R \rightarrow \infty$.
In the following we will construct an interacting theory
satisfying both criteria.
The cost we have to pay to meet these criteria is a nonlocal treatment 
of the compactified dimension, 
as we will see in the following sections.
Such a description may seem exotic, 
but it might be justified 
in view of the special role played by the compactified direction
in defining M-theory as the strong coupling limit of type \IIA string theory.

We organize this paper as follows.  
In section 2, we define the non-Abelian gauge transformation 
for an anti-symmetric two-form.
A characteristic feature is to introduce separate treatments
for zero-mode and non-zero mode (KK mode) in the compactified direction.
In section 3, after a brief review of Lagrangian formulation of self-dual two-form, 
we explain how to modify the equation of motion 
to include the non-Abelian gauge symmetry.
In section 4, we proposes an action which produces 
the equation of motion.  
We also describe how D4 brane action can be
derived in the small radius limit.  
Finally in section 5,
we discuss some related issues such as the recovery of Lorentz symmetry 
in large R limit and comparison with the D4-brane approach.

\section{Gauge Symmetry}

\subsection{Gauge Transformations}
\label{GT}

As explained above,
here we consider the case where the world volume of M5 is
$\mathbf{R}^5\times S^1$ where $S^1$ is a circle of radius $R$ with a
coordinate $x^5\sim x^5+2\pi R$.
We will use the notation
that the superscript ``$\z$'' represents zero modes
and ``(KK)'' represents Kaluza-Klein (non-zero) modes.
For an arbitrary field $\Phi$,
we have the decomposition
\be
\Phi = \Phi^{\z} + \Phi^{\nz}.
\ee
Obviously,
\be
\del_5 \Phi^{\z} = 0, \qquad
\del_5 \Phi = \del_5 \Phi^{\nz}.
\ee
We can define a non-local operator $\del_5^{-1}$
on the space of KK modes,
so that, for instance, $\del_5^{-1}\Phi^{\nz}$ is well defined.

After a lot of trial and error,
we find a consistent non-Abelian generalization of
the gauge transformation for a 2-form gauge potential $B_{\mu\nu}$
as
\bea
\d B_{i5} &=& [D_i, \Lam_5] - \del_5 \Lam_i + g[B_{i5}^{\nz}, \Lam_5^{\z}],
\label{dBi5} \\
\d B_{ij} &=& [D_i, \Lam_j] - [D_j, \Lam_i]
+ g[B_{ij}, \Lam_5^{\z}] - g[F_{ij}, \del_5^{-1}\Lam_5^{\nz}],
\label{dBij}
\eea
where $i, j = 0, 1, 2, 3, 4$.
The covariant derivative is defined in terms of the zero-mode:
\be
D_i \equiv \del_i + gB_{i5}^{\z}.
\ee
The parameter $g$ is the coupling constant for
the 2-form gauge interaction.
The field strength is
\be
F_{ij} \equiv g^{-1}[D_i, D_j]
= \del_i B_{j5}^{\z} - \del_j B_{i5}^{\z} + g[B_{i5}^{\z}, B_{j5}^{\z}].
\ee
Eqs. (\ref{dBi5}) and (\ref{dBij}) can be decomposed into
their zero modes and KK modes as
\bea
\d B_{i5}^{\z} &=& [D_i, \Lam_5^{\z}],
\label{dBi50} \\
\d B_{i5}^{\nz} &=& [D_i, \Lam_5^{\nz}] - \del_5 \Lam_i^{\nz} + g[B_{i5}^{\nz}, \Lam_5^{\z}],
\label{dBi5KK} \\
\d B_{ij}^{\z} &=& [D_i, \Lam_j^{\z}] - [D_j, \Lam_i^{\z}] + g[B_{ij}^{\z}, \Lam_5^{\z} ],
\label{dBij0} \\
\d B_{ij}^{\nz} &=& [D_i, \Lam_j^{\nz}] - [D_j, \Lam_i^{\nz}]
+ g[B_{ij}^{\nz}, \Lam_5^{\z} ] - g[F_{ij}, \del_5^{-1}\Lam_5^{\nz}].
\label{dBijKK}
\eea
All quantities $B_{i5}, B_{ij}, \Lam_i, \Lam_5$
take values in a Lie algebra $h$.
The last term in (\ref{dBij}) (or (\ref{dBijKK}))
is the only explicit nonlocality in these expressions.
But in fact there is additional nonlocality introduced 
by how the gauge transformations are defined separately 
for the zero modes and KK modes.

If $h$ is abelian, the transformations (\ref{dBi5}) and (\ref{dBij})
are equivalent to the conventional gauge transformation of two-form
gauge potential.  For our non-abelian generalization,
the gauge transformation laws explicitly distinguish
the zero modes from the KK modes
and we treat them as if they are independent fields.
All the commutators involve at most one KK mode.
There is no term of the form $[B^{\nz}, \Lambda^{\nz}]$
in the transformation laws.
Some of the physical meanings of these peculiar features 
will be explained in Sec.\;\ref{interpret}.

Because of these choices,
the gauge transformations (\ref{dBi50}, \ref{dBij0}) are closed
by zero-mode fields/gauge parameters.  
While they resemble the gauge transformation of
non-abelian gerbe (\ref{nag1}, \ref{nag2}) if one replaces $B^{\z}_{i5}$ by $A_i$, 
these are different since the $\Lambda$ term in (\ref{nag1}) 
and the nonlinear term in (\ref{nag2}) are absent in (\ref{dBi50}).  
In a sense, the transformation by the vector gauge parameter
$\Lambda^{(0)}_i$ is abelian (no $\Lambda^2$ term) and the noncommutativity comes
in through the transformation by $\Lambda^{(0)}_5$.
Our choice is more useful to realize the self-dual field
after the transformations of KK modes (\ref{dBi5KK}, \ref{dBijKK}) are included.

There are 6 gauge transformation parameters $\Lam_i, \Lam_5$,
but only 5 of the KK modes are independent
because the gauge transformation parameters are defined up to
the transformation
\be
\d \Lam_i^{\nz} = [D_i, \lam^{\nz}], \qquad
\d \Lam_5^{\nz} = \del_5 \lam^{\nz}.
\ee
This ``gauge symmetry of gauge symmetry'' is crucial for the gauge symmetry
to be justified as a deformation of the Abelain gauge symmetry.
We can use this redundancy to ``gauge away'' $\Lam_5^{\nz}$.
That is, the gauge transformation rules (\ref{dBi5}, \ref{dBij})
are equivalent to
\bea
\d B_{i5} &=& - \del_5 \Lam'_i + g[B_{i5}, \Lam_5^{\z}],
\label{dBi5-1} \\
\d B_{ij} &=& [D_i, \Lam'_j] - [D_j, \Lam'_i] + g[B_{ij}, \Lam_5^{\z}],
\label{dBij-1}
\eea
where
\be
\Lam'_i \equiv \Lam_i - [D_i, \del_5^{-1}\Lam_5^{\nz}].
\ee
However, the zero mode $\Lam_5^{\z}$ can not be gauged away.
Note that the only nonlocal term in the gauge transformation
(the last term in (\ref{dBij}))
is gauged away through this change of variables.

The 3-form field strengths are defined as
\bea
H_{ij5}^{\z} &\equiv& F_{ij} \equiv g^{-1}[D_i, D_j],
\label{Hij50} \\
H_{ij5}^{\nz} &\equiv& [D_i, B_{j5}^{\nz}] - [D_j, B_{i5}^{\nz}] + \del_5 B_{ij},
\label{Hij5KK} \\
H_{ijk}^{\z} &\equiv& [D_i, B_{jk}^{\z}] + [D_j, B_{ki}^{\z}] + [D_k, B_{ij}^{\z}],
\label{Hijk0} \\
H_{ijk}^{\nz} &\equiv& [D_i, B_{jk}^{\nz}] + [D_j, B_{ki}^{\nz}] + [D_k, B_{ij}^{\nz}]
\nn \\
&&
+ g[F_{ij}, \del_5^{-1}B_{k5}^{\nz}]
+ g[F_{jk}, \del_5^{-1}B_{i5}^{\nz}]
+ g[F_{ki}, \del_5^{-1}B_{j5}^{\nz}].
\label{HijkKK}
\eea
They satisfy the generalized Jacobi identities
\bea
\sum_{(3)} [D_i, H_{jk5}^{\z}] &=& 0, \\
\sum_{(3)} [D_i, H_{jk5}^{\nz}] &=& \del_5 H_{ijk}^{\nz}, \\
\sum_{(4)} [D_i, H_{jkl}^{\z}] &=& 0, \\
\sum_{(4)} [D_i, H_{jkl}^{\nz}] &=& g \sum_{(6)} [H_{ij5}^{\z}, \del_5^{-1}H_{kl5}^{\nz}],
\eea
where $\sum_{(n)}$ represents a sum over $n$ terms
that totally antisymmetrizes all the indices.
The field strength transforms as
\bea
\d H_{ij5}^{\z} &=& g[H_{ij5}^{\z}, \Lam_5^{\z}], \\
\d H_{ij5}^{\nz} &=& g[H_{ij5}^{\nz}, \Lam_5^{\z}], \\
\d H_{ijk}^{\z} &=& g[H_{ijk}^{\z}, \Lam_5^{\z}]
+ g[H_{ij5}^{\z}, \Lam_k^{\z}] + g[H_{jk5}^{\z}, \Lam_i^{\z}] + g[H_{ki5}^{\z}, \Lam_j^{\z}],
\label{dHijk0} \\
\d H_{ijk}^{\nz} &=& g[H_{ijk}^{\nz}, \Lam_5^{\z}].
\eea

The components $B_{i5}^{\nz}$ can be gauged away
using the gauge transformations parametrized by $\Lam_i^{\nz}$.
In this gauge, $B_{i5}^{\nz} = 0$,
we have $H_{ij5}^{\nz} = \del_5 B_{ij}^{\nz}$.
This motivates us to define
\be
\hat{B}_{ij}^{\nz} \equiv \del_5^{-1}H_{ij5}^{\nz},
\ee
which transforms covariantly as
\be
\d \hat{B}_{ij}^{\nz} = g[\hat{B}_{ij}^{\nz}, \Lam_5^{\z}],
\ee
and then (\ref{Hij5KK}) and (\ref{HijkKK}) are equivalent to
\bea
H^{\nz}_{ij5} &\equiv& \del_5\hat{B}_{ij}^{\nz}, \\
H^{\nz}_{ijk} &\equiv&
[D_i, \hat{B}_{jk}^{\nz}] + [D_j, \hat{B}_{ki}^{\nz}] + [D_k, \hat{B}_{ij}^{\nz}].
\eea

The algebra of gauge transformations is closed and given by
\be
[\d, \d'] = \d'',
\ee
with
\bea
\Lam''_5{}^{\z} &=& g[\Lam_5^{\z}, \Lam'_5{}^{\z}], \\
\Lam''_5{}^{\nz} &=& g[\Lam_5^{\z}, \Lam'_5{}^{\nz}] - g[\Lam'_5{}^{\z}, \Lam_5^{\nz}], \\
\Lam''_i &=& g[\Lam_5^{\z}, \Lam'_i] - g[\Lam'_5{}^{\z}, \Lam_i].
\eea

\subsection{Coupling to Antisymmetric Tensors}

Apart from the application to multiple M5-branes,
let us also consider applications to non-Abelian 2-form gauge theories
which are not self dual.
A potential problem is that the transformation of $H_{ijk}^{\z}$ (\ref{dHijk0})
is different from the usual covariant form like other components of $H$.
If we couple other tensor fields to the gauge field,
they will have to transform in a similar way.
A straightforward generalization of the transformation laws for $H$
leads to the definition of gauge transformations of 
a totally antisymmetrized tensor field $\phi_{\mu_1\cdots \mu_n}$
($n \leq 6$),
which can be decomposed into a multiplet
$(\phi_{i_1\cdots i_{n-1} 5}^{\z}, \phi_{i_1 \cdots i_{n-1} 5}^{\nz},
\phi_{i_1\cdots i_n}^{\z}, \phi_{i_1\cdots i_n}^{\nz})$,
to be
\bea
\d \phi_{i_1\cdots i_{n-1} 5}^{\z} &=&
g[\phi_{i_1\cdots i_{n-1} 5}^{\z}, \Lam_5^{\z}], \\
\d \phi_{i_1 \cdots i_{n-1} 5}^{\nz} &=&
g[\phi_{i_1 \cdots i_{n-1} 5}^{\nz}, \Lam_5^{\z}], \\
\d \phi_{i_1\cdots i_n}^{\z} &=&
g[\phi_{i_1\cdots i_n}^{\z}, \Lam_5^{\z}]
+ g\sum_{(n)} [\phi_{i_1\cdots i_{n-1} 5}^{\z}, \Lam_{i_n}^{\z}],
\label{dphiijk0}\\
\d \phi_{i_1\cdots i_n}^{\nz} &=&
g[\phi_{i_1\cdots i_n}^{\nz}, \Lam_5^{\z}],
\eea
where $\sum_{(n)}$ represents a sum of $n$ terms
that totally antisymmetrizes all indices.

The transformation law (\ref{dphiijk0}) for the component $\phi_{i_1\cdots i_n}^{\z}$
is different from all other components.
It is defined to mimic the gauge transformation of $H_{ijk}^{\z}$.
We should check whether this complication will prevent us from
constructing a gauge field theory.
First,
products of these fields $\phi_{i_1\cdots i_n}^{\z}$ will also transform 
in the form of (\ref{dphiijk0})
when all indices are antisymmetrized on the products.
Secondly, 
the action of $D_i$ on $\phi_{i_1\cdots i_n}^{\z}$ does not transform covariantly,
but we can define a covariant exterior derivative for $\phi_{i_1\cdots i_n}^{\z}$ as
\be
({\cal D}\phi)_{i_1\cdots i_{n+1}}^{\z} \equiv
\sum_{(n+1)} [D_{i_{1}}, \phi_{i_2\cdots i_{n+1}}^{\z}]
- (-1)^n \sum_{((n+1)n/2)} [B_{i_1 i_2}^{\z}, \phi_{i_3\cdots i_{n+1} 5}^{\z}].
\label{covextder3}
\ee
(This expression is nontrivial only if $n \leq 5$.)
This covariant exterior derivative is indeed covariant, that is,
\be
\d ({\cal D}\phi)_{i_1\cdots i_{n+1}}^{\z} =
g[({\cal D}\phi)_{i_1\cdots i_{n+1}}^{\z}, \Lam_5^{\z}]
+ g\sum_{(n+1)} [({\cal D}\phi)_{i_1\cdots i_n 5}^{\z}, \Lam_{i_{n+1}}^{\z}],
\ee
where the exterior derivative of $\phi_{i_1 \cdots i_{n-1} 5}$ is defined by
\be
({\cal D}\phi)_{i_1\cdots i_n 5}^{\z}
= g\sum_{(n)} [D_{i_1}, \phi_{i_2\cdots i_n 5}^{\z}].
\ee
It seems possible to down covariant equations of motion 
using exterior derivatives and totally antisymmetrized tensors.

The real problem with the transformation law (\ref{dphiijk0}) 
lies in the definition of an invariant action.
For example, to define a Yang-Mills like theory,
the Lagrangian should look like
\be
\frac{1}{6}\mbox{Tr}(H_{ijk}^{\z}H^{\z}{}^{ijk} + 3H_{ij5}^{\z}H^{\z}{}^{ij5}
+ H_{ijk}^{\nz}H^{\nz}{}^{ijk} + 3H_{ij5}^{\nz}H^{\nz}{}^{ij5}).
\ee
Only the first term is not gauge invariant.
It is not clear how to modify the action to make it invariant.
Similarly it is hard to define the usual kinetic term for 
the components $\phi_{i_1\cdots i_n}^{\z}$ of a matter field.

In the following we will see that in a Lagrangian formulation
of the non-Abelian self-dual gauge theory in 6 dimensions,
we do not have to use the variables $B_{ij}^{\z}$ explicitly,
so the anomalous covariant transformation law
of $H_{ijk}^{\z}$ (\ref{dHijk0}) will never be used.
In fact we can simply define $H_{ijk}^{\z}$ to be the Hodge dual of $F_{ij}$,
so that its gauge transformation is the same as other components.
As a result the covariant transformation laws for matter fields
can be uniformly defined as
\be
\d \Phi = g[\Phi, \Lam_5^{\z}]
\ee
for all components of a matter field.

\section{Non-Abelianizing the Abelian Theory}
\label{Abelian}

The linearized Lorentz-covariant action
for an Abelian chiral 2-form potential is \cite{PST,Pasti}
\be
S = \frac{1}{4!} T_{M5}T_{M2}^{-2} \int d^6 x\; \left[
\frac{3}{(\del_{\rho}a\del^{\rho}a)}\del^{\mu}a
(H-\tilde{H})_{\mu\lam\sigma}(H-\tilde{H})^{\nu\lam\sigma}\del_{\nu}a
- H_{\mu\nu\lam}H^{\mu\nu\lam}
\right],
\ee
where
\be
H_{\mu\nu\lam} = \del_{\mu} B_{\nu\lam} + \del_{\nu} B_{\lam\mu} + \del_{\lam} B_{\mu\nu},
\ee
$\tilde{H}$ is the Hodge dual of $H$ and
$a$ is an auxiliary field.
In addition to the usual gauge symmetry for a 2-form potential
\be
\d B_{\mu\nu} = \del_{\mu}\Lam_{\nu} - \del_{\nu}\Lam_{\mu},
\label{dBdL}
\ee
it is invariant under two gauge transformations
\be
\d B_{\mu\nu} = (\del_{\mu}a) \Phi_{\nu}(x) - (\del_{\nu}a) \Phi_{\mu}(x),
\qquad
\d a = 0,
\label{dBPhi-0}
\ee
and
\be
\d B_{\mu\nu} = \frac{\varphi(x)}{(\del a)^2}(H-\tilde{H})_{\mu\nu\rho}\del^{\rho}a,
\qquad
\d a = \varphi(x).
\label{da}
\ee
Using the gauge symmetry (\ref{da}),
one can impose the gauge fixing condition
\be
a = x^5,
\ee
so that the action becomes
\be
S = \frac{1}{4} T_{M5}T_{M2}^{-2} \int d^6 x \; \left(
\frac{1}{6}\eps^{ijklm}H_{ijk}\left[
H_{lm5} + \frac{1}{6}\eps_{lmnpq} H^{npq}
\right]
\right).
\ee
The 6 dimensional Lorentz symmetry is still preserved but
with a modified transformation law for a boost parametrized by $v_k$ as
\be
\d B_{ij} = x^5 v_k \del^k B_{ij} - x^k v_k \del^5 B_{ij} - x^k v_k (H-\tilde{H})_{ij5}.
\ee
The gauge transformation (\ref{dBPhi-0}) reduces in this gauge to
\be
\d B_{i5} = \Phi_i.
\label{dBPhi}
\ee

Now we consider the compactification of the Abelian theory on a circle of radius $R$ along $x^5$.
All fields can be decomposed into their zero modes and KK modes,
and the action becomes
\be
S = S^{\z} + S^{\nz},
\ee
where
\bea
S^{\z} &=& \frac{2\pi R}{12} T_{M5}T_{M2}^{-2} \int d^5 x \; H^{\z}_{ijk}H^{\z}{}^{ijk},
\\
S^{\nz} &=& \frac{1}{4} T_{M5}T_{M2}^{-2} \int d^6 x \; \left(
\frac{1}{6}\eps^{ijklm}H^{\nz}_{ijk}\left[
H^{\nz}_{lm5} + \frac{1}{6}\eps_{lmnpq} H^{\nz}{}^{npq}
\right]
\right).
\label{SKKA}
\eea
The zero modes $B_{ij}^{\z}$ are 5 dimensional 2-form potential,
and we can carry out the standard procedure of electric-magnetic duality
for $S^{\z}$ to get an action for the dual 1-form potential
\be
S^{\z}_{dual} = \frac{2\pi R}{4} T_{M5}T_{M2}^{-2} \int d^5 x \; F_{ij} F^{ij},
\label{S0A}
\ee
where $F_{ij} = H^{\z}_{ij5}$ is the field strength of the dual 1-form potential $B_{i5}^{\z}$.

Let us check that the equations of motion derived from the new action
$S^{\z}_{dual} + S^{\nz}$ lead to configurations satisfying self-duality conditions.
For the zero modes,
the equation of motion derived from the action $S^{\z}_{dual}$ is
\be
\del^j F_{ij} = 0.
\label{dF=0}
\ee
Defining a 3-form field $H$ by
\be
H_{ijk}^{\z} = \frac{1}{2} \eps_{ijklm} F^{lm},
\label{HF}
\ee
we see that, due to the equation of motion (\ref{dF=0}),
a 2-form potential $B^{\z}$ exists locally
such that $H^{\z} = dB^{\z}$.
Since $F$ also satisfies the Jacobi identity $dF = 0$,
we find
\be
\del^k H_{ijk}^{\z} = 0.
\label{dH=0}
\ee
Note that (\ref{HF}) is identical to the self-duality condition
for the zero modes
\be
H_{ijk}^{\z} = \frac{1}{2}\eps_{ijklm} H^{\z}{}^{lm5}
\ee
by identifying $A_i$ with $B_{i5}$.
Hence we see that the zero modes of the self-dual gauge field
can be simply described by the Maxwell action $S^{\z}_{dual}$.

It is natural to non-Abelianize the equation of motion (\ref{dF=0})
for the zero modes by
\be
[D^j, F_{ij}] = 0 + \cdots,
\label{SD-z}
\ee
up to additional covariant terms that vanish when the Lie algebra $h$ is Abelian.
In the next section we will derive the complete equation
from an action principle.

For the non-Abelian theory described in Sec.\;\ref{GT},
we could also have defined $H^{\z}_{ijk}$ simply as
the Hodge dual of $F_{ij}$,
hence it is not necessary to introduce the components $H^{\z}_{ijk}$
which has the unusual transformation law (\ref{dHijk0}).
The transformation of $F_{ij}$ would then imply that $H^{\z}$,
defined as the Hodge dual of $F_{ij}$, 
transforms simply as
\be
\d H^{\z}_{ijk} = [H^{\z}_{ijk}, \Lambda^{\z}_5].
\ee
This would also lead us to redefine the transformation laws
of matter fields as
\be
\d \phi = [\phi, \Lambda^{\z}_5]
\ee
for all components of $\phi$.

For the KK modes, the equations of motion derived from varying $S^{\nz}$ is
\be
\eps^{ijklm} \del_k \left(H^{\nz}_{lm5} + \frac{1}{6}\eps_{lmnpq} H^{\nz npq}\right)
= 0.
\label{dHH=0}
\ee
This implies that
\be
\eps^{ijklm}\left(H^{\nz}_{lm5} + \frac{1}{6}\eps_{lmnpq} H^{\nz npq}\right)
= \eps^{ijklm}\Phi_{\nz lm}
\label{dHHPhi}
\ee
for some tensor $\Phi_{lm}^{\nz}$ satisfying
\be
\eps^{ijklm}\del_k \Phi^{\nz}_{lm} = 0.
\label{cond-Phi}
\ee
We can redefine $B_{lm}^{\nz}$ by a shift
\footnote{
As (\ref{cond-Phi}) implies that
$\Phi_{lm}^{\nz} = \del_l \Phi_m^{\nz} - \del_m \Phi_l^{\nz}$
for some vector field $\Phi_l^{\nz}$,
in the Abelian theory
the effect of shifting $B_{lm}^{\nz}$ is equivalent to
the shift of $B_{l5}^{\nz}$ in the gauge symmetry (\ref{dBPhi}), 
up to the usual gauge transformation (\ref{dBdL}).
}
\be
B_{lm}^{\nz} \rightarrow B_{lm}^{' \nz} \equiv B_{lm}^{\nz} + \del_5^{-1} \Phi_{lm}^{\nz}
\ee
such that, due to (\ref{cond-Phi}),
\bea
H^{\nz}_{lm5} &\rightarrow& H^{' \nz}_{lm5} \equiv H^{\nz}_{lm5} + \Phi^{\nz}_{lm}, \\
\eps^{lmnpq} H^{\nz}_{npq} &\rightarrow& \eps^{lmnpq} H^{' \nz}_{npq}
\equiv \eps^{lmnpq} (H^{\nz}_{npq}
+ 3\del_n \Phi_{pq}^{\nz}) = \eps^{lmnpq} H^{\nz}_{npq}.
\eea
As a result, (\ref{dHHPhi}) is turned into the self-duality condition
\be
H^{\nz}_{lm5} = -\frac{1}{6}\eps_{lmnpq} H^{\nz npq}.
\label{SD-nz-direct2}
\ee

Let us define the non-Abelian counterpart of (\ref{dHH=0}) as
\be
\eps^{ijklm} \left[D_k, \left(H^{\nz}_{lm5}
+ \frac{1}{6} \eps_{lmnpq} H^{\nz npq}\right)\right]
= 0.
\label{SD-nz}
\ee
This implies that
\be
\eps^{ijklm} \left(H^{\nz}_{lm5} + \frac{1}{6} \eps_{lmnpq} H^{\nz npq}\right)
= \eps^{ijklm} \Phi^{\nz}_{lm},
\ee
where $\Phi^{\nz}_{lm}$ satisfies
\be
\eps^{ijklm} [D_k, \Phi^{\nz}_{lm}] = 0.
\label{constraint-Phi}
\ee
This again can be absorbed into a shift of $B_{lm}^{\nz}$
\be
B_{lm}^{\nz} \rightarrow B_{lm}^{' \nz} \equiv B_{lm}^{\nz} + \del_5^{-1} \Phi_{lm}^{\nz},
\label{dBij=Phi}
\ee
so that the self-duality condition (\ref{SD-nz-direct2}) is arrived.

The transformation (\ref{dBij=Phi}) should also be viewed 
as a gauge transformation of the theory. 
The gauge transformation parameter $\Phi^{\nz}_{lm}$ has to transform covariantly 
under the transformation defined in Sec.\;\ref{GT} as
\footnote{
The meaning of the transformation of a gauge transformation parameter is this: 
$\Phi$ should be viewed as a function depending on 
the gauge potential $B_{i5}^{\z}$ as well as some free parameters 
corresponding to integration constants 
when we solve the constraint (\ref{constraint-Phi}).
The transformation of $B_{i5}^{\z}$ induces a transformation of $\Phi$.
}
\be
\d \Phi^{\nz}_{lm} = [\Phi^{\nz}_{lm}, \Lam_5^{\z}],
\ee
because the constraint (\ref{constraint-Phi}) is covariant.
It can then be checked that (\ref{dBij=Phi}) commutes with 
the gauge transformation (\ref{dBi5}, \ref{dBij}) defined in Sec.\;\ref{GT}.
Our task in the next section is to give an action that would lead to
the non-Abelian equations of motion (\ref{SD-z}) and (\ref{SD-nz}).

\section{Action}
\label{action}

Let us consider the following action for the non-Abelian chiral 2-form potential
\be
S = S^{\z} + S^{\nz},
\ee
where
\bea
S^{\z} &=& \frac{2\pi R}{4} T_{M5}T_{M2}^{-2} \int d^5 x \; \mbox{Tr}(F_{ij}F^{ij}),
\label{S0-1} \\
S^{\nz} &=& \frac{1}{4} T_{M5}T_{M2}^{-2} \int d^6 x \; \mbox{Tr}\left(
\frac{1}{6} \eps^{ijklm}H_{ijk}^{\nz}\left[
H_{lm5}^{\nz} + \frac{1}{6}\eps_{lmnpq} H^{\nz}{}^{npq}
\right]
\right).
\label{SKK-1}
\eea
This invariant action
is a straightforward generalization of the action (\ref{SKKA}) and (\ref{S0A})
for the Abelian theory.

For small $R$,
the M5-branes should be approximated by D4-branes in type \IIA theory,
so $S^{\z}$ should be identified the Yang-Mills theory for multiple D4-branes
\be
S^{\z} = \frac{1}{4} T_{D4} T_s^{-2} \int d^5 x \; \mbox{Tr}(f_{ij}f^{ij}),
\label{S0-2}
\ee
where the field strength $f_{ij}$ for multiple D4-branes is
\be
f_{ij} \equiv [\del_i + A_i, \del_j + A_j] = \del_i A_j - \del_j A_i + [A_i, A_j].
\ee
It is known that the gauge potential $A$ in D4-brane theory
is related to the gauge potential $B$ in M5-brane theory via the relation
\be
A_i = 2\pi R B_{i5}^{\z}.
\ee
Plugging in the values of the parameters involved,
\be
T_{M5} = \frac{1}{2\pi} T_{M2}^2, \qquad
T_{D4} = \frac{1}{(2\pi)^4 g_s \ell_s^5}, \qquad
T_s = \frac{1}{2\pi \ell_s^2}, \qquad
R = g_s \ell_s,
\ee
we find that the coupling constant should be given by
\be
g = 2\pi R.
\ee
This factor can also be obtained by demanding that
the soliton solutions which resemble instantons in the spatial 4 dimensions
have momentum equal to $n/R$ for some integer $n$ in the $x^5$ direction.

Notice that the overall factor of $2\pi R$ due to the integration over $x^5$
in (\ref{S0-1}) is multiplied by a factor of $1/g^2$
for $g$ is the Yang-Mills coupling for the zero mode field strength $F_{ij}$,
giving an overall factor of $1/R$ in (\ref{S0-2}),
in agreement with the requirement of conformal symmetry in 6 dimensions.
Witten's argument \cite{Witten}
mentioned in the introduction around (\ref{1/R}) and (\ref{R})
that M5-brane action does not exist is resolved 
by allowing 
the coupling of a 6 dimensional theory to
depend on the compactification radius $R$.
Normally the coupling constant of an interacting field theory is
independent of whether the space is compactified.
Our strategy is to define a 6 dimensional field theory as
the decompactification limit of a compactified theory,
and the coupling depends on the compactification radius.
In some sense, the coupling constant $g$
is not really the coupling of the decompactified theory,
which is a conformal field theory without free parameter.
Witten's argument should be understood as the non-existence
of another formulation of the 6 dimensional theory
which does not refer to the compactification radius.
In other words, our model may be as good as it gets
if we want to describe multiple M5-branes with an action.

Assuming that we will be able to show in future works that
a well defined theory does exist in uncompactified 6 dimensional spacetime
as the decompactification limit of our model,
one would still wonder how such a theory
can be fully Lorentz invariant,
while its definition involves the choice of a special direction.
We will discuss more on this point in the next section.

The variation of $S^{\z}$ leads to Yang-Mills equations,
which can be interpreted as the self-dual equation for the zero modes.
The full equation of motion for the zero modes $B_{i5}^{\z}$
should also include variations of $S^{\nz}$,
which modifies the Yang-Mills equation by commutators
that vanish in the Abelian case.

Contrary to the proposal of \cite{Douglas,LPS},
we have an explicit appearance of the KK modes through the action $S^{\nz}$ (\ref{SKK-1}).
A useful feature of $S^{\nz}$ is that it depends on
$B_{i5}^{\nz}$ and $B_{ij}$ only through $\hat{B}_{ij}^{\nz}$.
Therefore, although $S^{\nz}$ depends on both $B_{i5}^{\nz}$ and $B_{ij}^{\nz}$
(unlike the Abelian case where the action is independent of $B_{i5}$),
we only need to consider the variation of $\hat{B}_{ij}^{\nz}$.
Variation of the action $S^{\nz}$ with respect to $\hat{B}_{ij}^{\nz}$
leads to the equation of motion (\ref{SD-nz}),
which is equivalent to the self-duality condition (\ref{SD-nz-direct2})
via a shift in $B_{lm}^{\nz}$,
as we explained in the previous section.
Explicitly,
the equations of motion are
\bea
&[D^j, F_{ij}] = \frac{1}{2} \int_0^{2\pi R} dx^5 \;
\left[\hat{B}_{jk}, \left(H^{\nz}{}^{ijk} - \frac{1}{4} \eps^{ijklm} H^{\nz}_{lm5}\right)\right],
\label{EOM1} \\
&\eps^{ijklm} \left[D_k, \left(H^{\nz}_{lm5}
+ \frac{1}{6} \eps_{lmnpq} H^{\nz npq}\right)\right] = 0.
\label{EOM2}
\eea

\section{Discussions}
\label{interpret}

\subsection{$R \rightarrow \infty$}

The reader may find it strange that in the gauge transformation laws
(\ref{dBi5KK}) and (\ref{dBijKK}) we have avoided commutators
involving two KK modes, e.g. terms of the form $[B^{\nz}, \Lambda^{\nz}]$.
Correspondingly, there is no term of the form $[B^{\nz}, B^{\nz}]$
in the equations of motion.
All gauge interactions are mediated via zero modes.
Here is our interpretation.
In the limit $R \rightarrow \infty$,
the Fourier expansion of a field approaches to the Fourier transform
\be
\Phi(x^5) = \sum_n \Phi_n e^{in x^5/R}
\qquad \longrightarrow \qquad
\Phi(x^5) = \int \frac{dk_5}{2\pi} \; \tilde{\Phi}(k_5) e^{ik_5 x^5}.
\ee
The coefficients $\Phi_n$ approach to $\tilde{\Phi}(k_5)$ as
\be
\tilde{\Phi}(k_5) = 2\pi R \Phi_n \qquad
(k_5 = n/R).
\label{PhiPhi}
\ee
According to this expression,
the value of a specific Fourier mode $\Phi_n$ must approach to zero
in the limit $R \rightarrow \infty$.
In particular, the amplitude of the zero mode approaches to zero.
While all interactions are mediated via the zero mode,
this does not imply that there is no interaction in the infinite $R$ limit,
because the coupling $g = 2\pi R \rightarrow \infty$.
The product of the amplitude of the zero mode with the coupling
is actually kept finite in the limit.

In the limit $R \rightarrow \infty$,
the KK modes $B_{\mu\nu}^{\nz}$ should be identified with the 2-form potential
in uncompactified 6 dimensional spacetime.
In uncompactified space,
the constant part of $B_{\mu\nu}$ is not an observable,
hence physically the KK modes $B_{\mu\nu}^{\nz}$ do not
miss any physical information a 2-form potential can carry.
The zero modes $B_{\mu\nu}^{\z}$ approach to zero but
a new field $A_i$ replacing $2\pi RB_{i5}^{\z}$ survives the large $R$ limit.
The field $A_i$ can not be viewed as part of the 2-form potential,
in the sense that, due to the infinite scaling of $B_{i5}^{\z}$ by $R$,
it can not be combined with $B_{\mu\nu}^{\nz}$
in a Lorentz covariant way to form a new tensor in 6 dimensions.
Rather it should be understood as the 1-form needed
to define gerbes (or some similar geometrical structure)
together with the 2-form potential.
However this does not increase the physical degrees of freedom of the 6 dimensional theory
in the sense that the number of physical degrees of freedom in the 5 dimensional field $A_i$
is negligible compared with that of a 6 dimensional field.

The fact that gauge transformation laws do not have
terms of the form $[B^{\nz}, \Lambda^{\nz}]$,
and the fact that the equations of motion
do not have terms of the form $[B^{\nz}, B^{\nz}]$,
are both telling us that our model is linearized with respect to the 2-form potential.
No self-interaction of the 2-form potential is present,
and all interactions are mediated by the 1-form potential $A_i$.

As the decompactification limit $R \rightarrow \infty$
is also the strong coupling limit $g \rightarrow \infty$,
we do not expect the classical equations of motion (\ref{EOM1})--(\ref{EOM2})
to give a good approximation of the quantum theory.
We leave the problem of finding the decompactification limit of our theory
for future study.

\subsection{Momentum and Instanton}
\label{proposal}

The interpretation above allows us to understand some puzzles
about the proposal of \cite{Douglas,LPS} that
the 5 dimensional D4-brane theory is already sufficient
to describe the 6 dimensional M5-brane system even for finite $R$.
In their proposal,
the momentum $p_5$ in the 5-th (compactified) direction is
represented by the ``instanton'' number on the 4 spatial dimensions.
The first problem with this interpretaion is that,
in the phase when $U(N)$ symmetry is broken to $U(1)^{N}$,
there is no instanton solution.
But physically this corresponds to having M5-brane well separated
from each other,
and they should still be allowed to have nonzero $p_5$.
This problem does not exist in our model.
In our model $p_5$ is carried by the KK modes
when the Lie algebra of the gauge symmetry is Abelian.
Furthermore, the Abelian case of our model is already known to
be equivalent to a 6 dimensional theory which has
the full Lorentz symmetry in the large $R$ limit.

The second problem of the proposal in \cite{Douglas,LPS} is that
the instanton number only gives the total value of $p_5$ of a state,
but it is unclear how to specify the distribution of $p_5$ over
different physical degrees of freedom.
For example, the state with $m$ units of $p_5$ contributed from
the scalar field $X^1$ and $n$ units of $p_5$ from $X^2$
cannot be distinguished from the state with the numbers $m$ and $n$ switched.
On the other hand, in our model, the instanton number of the 1-form
\be
A_i \equiv RB_{i5}^{\z}
\ee
should only be interpreted as the value of $p_5$ of the field $A_i$.
(In other words, the so-called ``zero-modes'' $B_{i5}^{\z}$ can still carry nonzero $p_5$.
The 5-th momentum of the 2-form potential is manifest as the KK mode index.)
The scalar fields $X^I$ and the fermions $\Psi$,
when they are introduced into our model,
would have their own KK modes to specify their $p_5$ contribution.
There is no ambiguity in the momentum carrier for a given instanton number.

The reader may wonder whether it is redundant or over-counting
for $A_i$ to be able to carry nontrivial $p_5$.
After all, $A_i$ is just $B_{i5}^{\z}$ rescaled.
Has not the KK modes $B_{i5}^{\nz}$ already taken care of
the contribution of $B_{i5}$ to $p_5$?
How can a field carry momentum in the $x^5$-direction
if it has no fluctuation (e.g. propagating wave) in that direction?
The answer is simple.
It is well known in classical electrocmagnetism that
the simultaneous presence of constant electric and magnetic fields
carry momentum,
because the momentum density $p_i$ is proportional to $F^{0j}F_{ij}$.
In the temporal gauge $A_0 = 0$,
the conjugate momentum of $A_j$ is $\Pi^j \equiv \del_0 A^j$,
and the momentum density $p_i$ is proportional to
\be
F^{0j}F_{ij} = \Pi^j (\del_i A_j) - \Pi^j (\del_j A_i).
\label{FFAA}
\ee
The first term is the standard contribution of a field to momentum $p_i$.
We also have $(\del_0 \phi) (\del_i \phi)$ for a scalar field $\phi$.
But there is no analogue of the 2nd term for a scalar field.
It is possible for the 2nd term to be present because $A_i$ has a Lorentz index.
The zero mode of $A_i$ in the $x^i$ direction can also contribute to $p_i$
through this term.
Similarly, for a 3-form field strength $H$,
the momentum density of $p_5$ is proportional to $H_{0ab}H^{ab5}$
($a, b = 1, 2, 3, 4$),
which includes the zero mode contribution
\be
H_{0ab}^{\z}F^{ab} = \frac{1}{6}\eps_{0abcd5} F^{ab}F^{cd}
\ee
because $H^{\z}_{ab5} = F_{ab}$.
This is precisely the same expression as the instanton number density.
Note that there are also contributions to $p_5$ from the KK modes
$H_{0ab}^{\nz}H^{\nz}{}^{ab5}$ in addition to the zero mode contribution,
analogous to the first term in (\ref{FFAA}).

\subsection{Generalization to 3-form Gauge Potential}

In non-Abelianizing the gauge transformations of a 2-form potential,
the zero mode $\Lam_5^{\z}$ plays a special role.
We associate the special role played by $\Lam_5^{\z}$ 
to its topological nature:
while $\Lam_5^{\nz}$ can be ``gauged away'',
the zero mode $\Lam_5^{\z}$ corresponds to the Wilson line degree of freedom 
for the gauge transformation parameter $\Lambda$
along the circle in the $x^5$ direction.
In this section we generalize this association to construct
non-Abelian gauge transformations for a 3-form potential.

First we study the Abelian gauge theory for a 3-form potential
on the spacetime of $\mathbf{R}^d\times T^2$.
Let the torus $T^2$ extend in the directions of $x^1$ and $x^2$.
We can decompose a field $\Phi$ as 
\be
\Phi = \Phi^{\z} + \Phi^{\nz},
\ee
where the zero mode $\Phi^{\z}$ has no dependence on $T^2$
\be
\del_a \Phi^{\z} = 0,
\ee
and the KK mode $\Phi^{\nz}$ can be obtained from $\Phi$ as 
\be
\Phi^{\nz} = \Box^{-1}\Box\Phi,
\ee
where 
\be
\Box \equiv \del^a \del_a 
\qquad
(a = 1, 2).
\ee

The Abelian gauge transformations of a 3-form potential $B$ are given by
\bea
\d B_{i12} &=& \del_i \Lam_{12} - \del_1 \Lam_{i2} + \del_2 \Lam_{i1}, \\
\d B_{ija} &=& \del_i \Lam_{ja} - \del_j \Lam_{ia} + \del_a \Lam_{ij}, \\
\d B_{ijk} &=& \del_i \Lam_{jk} + \del_j \Lam_{ki} + \del_k \Lam_{ij},
\eea
where $a = 1, 2$ and $i, j, k = 0, 3, 4, \cdots, (d+1)$.
There is redundancy in the gauge transformation parameters $\Lam_{ia}, \Lam_{ij}$ 
so that the gauge transformation laws are invariant under the transformation
\bea
\d \Lam_{12} &=& \del_1 \lam_2 - \del_2 \lam_1, \\
\d \Lam_{ia} &=& \del_i \lam_a - \del_a \lam_i, \\
\d \Lam_{ij} &=& \del_i \lam_j - \del_j \lam_i.
\eea
Apparently there is also a redundancy in using $\lam$ to parametrize
the redundancy in $\Lam$.
There are $(d+2)$ components in $\lam$,
but only $(d+1)$ of them are independent.
Using the redundancy of $\Lam$,
we can ``gauge away'' $(d+1)$ of the gauge transformation parameters.
For instance, we can set 
\be
\rho_i \equiv \del^a \Lam_{ia} = 0, 
\qquad
\Lam_{12} = 0,
\ee
and use the following gauge transformation parameters
\be
\xi_i \equiv \eps^{ab}\del_a \Lam_{ib}, 
\qquad
\Lam_{ij},
\ee
so that 
\be
\Lam_{ia}^{\nz} = -\eps^{ab}\Box^{-1}\del_b\xi_i,
\ee
and the gauge transformation laws become
\bea
\d B_{i12} &=& -\xi_i, 
\label{dBi12-0} \\
\d B_{ija} &=& -\eps^{ab}\Box^{-1}\del_b(\del_i\xi_j - \del_j\xi_i) + \del_a \Lam_{ij}, 
\label{dBija-0} \\
\d B_{ijk} &=& \del_i \Lam_{jk} + \del_j \Lam_{ki} + \del_k \Lam_{ij}.
\label{dBijk-0}
\eea
Viewing $\Lam_{ia}$ as $d$ copies of 1-form potentials on $T^2$,
the $\xi_i$'s are the corresponding field strengths, 
and so their integrals over $T^2$ are quantized.
It implies that $\xi_i^{\z}$ is quantized,
and so we have to set
\be
\xi_i^{\z} = 0
\ee
when we use $\xi_i$ as infinitesimal gauge transformation parameters.
In the following, we have $\xi_i = \xi_i^{\nz}$.

To retrieve from (\ref{dBi12-0})--(\ref{dBijk-0})
the original gauge transformation laws with redundancy,
one can simply carry out the replacement
\bea
\xi_i &\rightarrow& \xi_i - \del_i \Lam_{12}, \\
\Lam_{ij} &\rightarrow& \Lam_{ij} + \Box^{-1}(\del_i\rho_j - \del_j\rho_i).
\eea

On the torus $T^2$,
the gauge transformation parameter
$\Lam_{12}^{\z}$ corresponds to a Wilson surface degree of freedom
for the 2-form $\Lam$.
It should play the same role as $\Lam_5^{\z}$ in (\ref{dBi5}, \ref{dBij}).
To construct a consistent non-Abelian gauge transformation algebra for the 3-form potential,
we only need to consider transformation laws for the parameters $\xi_i^{\nz}$, $\Lam_{ij}$ and $\Lam_{12}^{\z}$.
In the end we get the full gauge transformation laws through the replacement
\bea
\xi_i^{\nz} &\rightarrow& \xi_i^{\nz} - [D_i, \Lam_{12}^{\nz}], \\
\Lam_{ij}^{\nz} &\rightarrow& \Lam_{ij}^{\nz} 
+ \Box^{-1}([D_i, \rho_j^{\nz}] - [D_j, \rho_i^{\nz}]),
\eea
where the covariant derivative $D_i$ should be defined as
\be
D_i = \del_i + B_{i12}^{\z},
\ee
and
\be
\xi_i^{\nz} \equiv \eps^{ab}\del_a\Lam_{ib}^{\nz}, 
\qquad
\rho_i^{\nz} \equiv \del^a\Lam_{ia}^{\nz}.
\ee
Here we have scaled $B_{i12}$ to absorb the coupling constant $g$,
which is expected to be given by the area $(2\pi)^2 R_1 R_2$ of the torus.

We define the non-Abelian gauge transformations as
\bea
\d B_{i12} &=& [D_i, \Lam_{12}^{\z}] - \xi_i + [B_{i12}^{\nz}, \Lam_{12}^{\z}], \\
\d B_{ija} &=& - \eps^{ab}\Box^{-1}\del_b\left([D_i, \xi_j]-[D_j, \xi_i]\right)
+ \del_a \Lam_{ij} \nn \\
&& + [D_i, \Lam_{ja}^{\z}] - [D_j, \Lam_{ia}^{\z}]
+ [B_{ija}. \Lam_{12}^{\z}], \\
\d B_{ijk} &=& [D_i, \Lam_{jk}] + [D_j, \Lam_{ki}] + [D_k, \Lam_{ij}]
+ [B_{ijk}, \Lam_{12}^{\z}].
\eea
The algebra of gauge transformations is closed
\be
[\d, \d'] = \d'',
\ee
with the parameters of $\d''$ given by
\bea
\Lam_{12}^{\z}{}'' &=& [\Lam_{12}^{\z}, \Lam_{12}^{\z}{}'], \\
\xi_i'' &=& [\xi_i, \Lam_{12}^{\z}{}'] - [\xi_i', \Lam_{12}^{\z}], \\
\Lam_{ij}'' &=& [\Lam_{ij}, \Lam_{12}^{\z}{}'] - [\Lam_{ij}', \Lam_{12}^{\z}].
\eea

The field strengths should be defined as
\bea
H_{ij12}^{\z} &=& [D_i, D_j], \\
H_{ij12}^{\nz} &=& [D_i, B_{j12}^{\nz}]-[D_j, B_{i12}^{\nz}]+\del_4 B_{ij5}^{\nz}-\del_5 B_{ij4}^{\nz}, \\
H_{ijka} &=& [D_i, B_{jka}] + [D_j, B_{kia}] + [D_k, B_{ija}] - \del_a B_{ijk} \nn \\
&& - \eps^{ab} \Box^{-1}\del_b\left( [F_{ij}, B_{k12}^{\nz}] + [F_{jk}, B_{i12}^{\nz}] + [F_{ki}, B_{j12}^{\nz}] \right), \\
H_{ijkl} &=& \sum_{(4)} [D_i, B_{jkl}] - \sum_{(6)} [F_{ij}, \beta_{kl}],
\eea
where
\be
\beta_{ij} \equiv \Box^{-1}\del^a B_{ija},
\ee
so that all the field strength components transform as
\bea
\d H_{ij12} &=& [H_{ij12}, \Lam_{12}^{\z}], \\
\d H_{ijka} &=& [H_{ijka}, \Lam_{12}^{\z}] + \sum_{(3)}[F_{ij}, \Lam_{ka}^{\z}], \\
\d H_{ijkl} &=& [H_{ijkl}, \Lam_{12}^{\z}] + \sum_{(6)}[F_{ij}, \Lam_{kl}^{\z}].
\eea
It may be possible to define a non-Abelian self-dual gauge theory for 
a 3-form potential in 8 dimensional Euclidean space. 
We leave the problem for future study.

Apparently, the same idea can be used to define 
a non-Abelian gauge symmetry for $p$-form potentials 
on $\mathbf{R}^d \times T^{p-1}$.

\subsection{Summary and Outlook}

In this paper,
we found a consistent, closed algebra of
non-Abelian gauge transformations for the 2-form potential
with 1-form gauge transformation parameters in 6 dimensions.
There is a gauge symmetry in the gauge transformations
parametrized by a 0-form.
The transformation law is nonlocal in the direction
which is compactified to a circle.

We also found an action which passes the two major tests for
it to be relevant to the M5-brane theory:
It is equivalent to
a Lorentz-invariant chiral 2-form theory in 6 dimensions
in the Abelian phase,
and it is equivalent to a Yang-Mills theory in 5 dimensions
in the limit $R\rightarrow 0$.

There are several additional tests this model has to pass
in order to prove that it is the correct theory for multiple M5-branes:
\begin{enumerate}
\item
In some sense there is
a well defined limit $R\rightarrow \infty$
in which the theory describes multiple M5-branes in uncompactified spacetime.
\item
6 dimensional Lorentz symmetry in the limit $R\rightarrow \infty$.
\item
The existence of a supersymmetric extension of the theory.
\end{enumerate}
In the above we have given persuasive arguments on these issues,
but we leave the full answers to these questions for future investigation.

\section*{Acknowledgment}

The authors thank Chong-Sun Chu, Kazuyuki Furuuchi, 
Hirotaka Irie, Sheng-Lan Ko, Tomohisa Takimi,
and Chi-Hsien Yeh for helpful discussions. 
Y. M. would like to thank the hospitality of people in Taipei 
during his stay in last September.
K. W. H. is grateful to Wung-Hong Huang for encouragements.
The work of P.-M. H. and K. W. H. is supported in part by
the National Science Council,
the National Center for Theoretical Sciences, 
and the LeCosPA Center at National Taiwan University. 
Y. M. is partially supported by
Grant-in-Aid (\#20540253) from 
the Japan Ministry of Education, Culture, Sports, Science and Technology.


\end{CJK}
\end{document}